\documentclass[a4paper,twoside]{article}

\usepackage{epsfig}
\usepackage{subcaption}
\usepackage{calc}
\usepackage{amssymb}
\usepackage{amstext}
\usepackage{amsmath}
\usepackage{amsthm}
\usepackage{multicol}
\usepackage{pslatex}
\usepackage{apalike}

\usepackage{multirow}
\usepackage[table,xcdraw]{xcolor}

\usepackage{SCITEPRESS}     

\pdfoutput=1

\begin{document}

\title{Proposal of a Novel Bug Bounty Implementation Using Gamification}

\author{\authorname{Jamie O'Hare\orcidAuthor{0000-0002-5847-6488}, Lynsay A. Shepherd\orcidAuthor{0000-0002-1082-1174}}
\affiliation{Division of Cyber Security, School of Design and Informatics, Abertay University, Dundee, United Kingdom}
\email{\{j.o'hare, lynsay.shepherd\}@abertay.ac.uk}}


\abstract{Despite significant popularity, the bug bounty process has remained broadly unchanged since its inception, with limited implementation of gamification aspects. Existing literature recognises that current methods generate intensive resource demands, and can encounter issues impacting program effectiveness. This paper proposes a novel bug bounty process aiming to alleviate resource demands and mitigate inherent issues. Through the additional crowdsourcing of report verification where fellow hackers perform vulnerability verification and reproduction, the client organisation can reduce overheads at the cost of rewarding more participants. The incorporation of gamification elements provides a substitute for monetary rewards, as well as presenting possible mitigation of bug bounty program effectiveness issues. Collectively, traits of the proposed process appear appropriate for resource and budget-constrained organisations - such Higher Education institutions.}
\onecolumn \maketitle \normalsize \setcounter{footnote}{0} \vfill

\section{\uppercase{Introduction}} \label{sec:introduction}

\noindent Traditionally, in-house teams, consulting penetration testers, or external good-natured researchers would identify and report security weaknesses to the accountable organisation. However, with large estates operating a diverse array of technologies, those responsible struggle to identify all possible vulnerabilities. As a result, a growing number of organisations are attempting to harness the power of crowdsourcing through the implementation of a bug bounty program to help combat this problem \cite{Walshe2020}. 

The past decade has seen the rapid adoption, development and maturity of bug bounty programs across a variety of organisations and sectors \cite{Walshe2020}. Organisations such as Google and Dropbox have adopted bug bounty programs, with the latter surpassing \$1 million in payouts on the HackerOne platform, while the former rewarded \$6.5 million in 2019, and over \$21 million since its inception \cite{Dropbox2020,Google2020}. Yet, innovation within the bug bounty sphere remains scarce. Minimal attention has been given to the further development of the crowdsourcing process mechanisms which underpin bug bounty programs, with limited studies only alluding to the potential for further reform \cite{Su2016Crowdsourcing}.

One area of potential innovation is the introduction of gamification. Current eco-systems exhibit limited implementation of any gamification aspects, and despite the developing sector, this design element has received scant attention in research concerning bug bounties. As such, it appears that no study examines the potential implementations and their possible effectiveness.

Through an exploration of the emerging role of gamification in an array of contexts, and an investigation of current bug bounty implementations, this paper proposes a new bug bounty implementation employing gamification aspects to provide a both a cybersecurity solution and an educational resource.

The remainder of this proposal follows four main sections. Section \ref{sec:background} provides an overview of existing bug bounty programs, and documents current gamification implementations in existing programs. Section \ref{sec:methodology} outlines the proposal of a novel bug bounty program utilising gamification, followed by a discussion of challenges and limitations in Section \ref{sec:discussion}. Finally, the paper ends with concluding remarks in Section \ref{sec:conclusion}.

\begin{figure}[th]
\centerline{\includegraphics[width=0.38\textwidth]{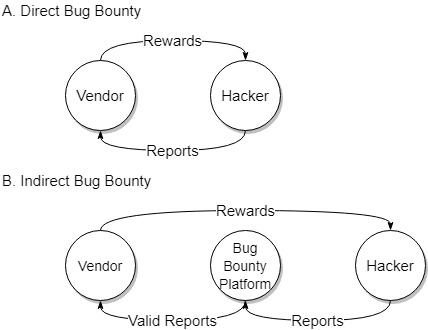}}
\caption{Variants of the bug bounty process. Adapted from \cite{ruohonen2018bug}.}
\label{figBB}
\end{figure}

\section{\uppercase{Background}}
\label{sec:background}

\subsection{Bug Bounty Programs}

Figure \ref{figBB} illustrates the most popular variants of the bug bounty process \cite{10.1145/3091478.3091517}. While in process A the external hacker corresponds directly to the vendor, process B introduces an intermediate platform. This intermediary can provide verification and triage before notification of the corresponding vendor, transferring the responsibility of gaining further detail from the hacker from the vendor to the platform, mitigating the concomitant administrative impact to the vendor. While a vendor can enrol on a bug bounty platform, they often pay additional fees for verification and triaging on top of bounty sums. Google and GitHub follow a direct bug bounty process, whereas many organisations use platforms such as HackerOne and Bugcrowd to facilitate the indirect process \cite{Google2020,GitHub2020,HackerOne2020,Bugcrowd2020}.

Typically, a hacker receives a monetary reward for a successful submission; however, for less critical vulnerabilities or conservative programs, branded vendor merchandise or kudos may be awarded \cite{laszka2018rules,ruohonen2018bug}. In some circumstances, a hacker can receive a further bonus award for a well-written explanatory report or a novel discovery \cite{malladi2019bug}. Furthermore, a subsection of ecosystems use reputation points and associated leaderboards to reward successful submissions \cite{GitHub2020}. These reputation points are often the criteria for admission to private programs \cite{HackerOne2020,Bugcrowd2020}.

While direct programs are often public, thus allowing for submissions from anyone, both major bug bounty ecosystems facilitate private programs, where only a subsection of hackers can see the program details and participate \cite{malladi2019bug}. The operation of a private program allows some organisations to test procedures before going public; however, some programs remain private for a significant amount of time or permanently. Consequently, these programs avoid some issues prevalent in public operations. 

\subsection{Bug Bounty Issues}
\label{bugbountyissues}

The emerging popularity of bug bounties has revealed systemic issues with existing crowdsourcing techniques. Previous studies document these issues and provide potential solutions; however, investigations examining their effectiveness remain limited.

A significant, prevailing issue in bug bounties is the high volume of low-quality submissions \cite{al2018friendly,zhao2016crowdsourced}. The insufficient report quality is a casualty of first-come first-served response to verified submissions, with hackers sometimes racing to submit a vulnerability. With many hackers concentrating on capitalising from their skills economically, many look to maximise the number of submissions rather than focusing on specific vulnerabilities \cite{hata2017understanding}. Potentially, this issue is encouraged with admission to private programs. Consequently, these problems can be a significant detriment to the effectiveness of a program, often impeding the verification and triaging processes \cite{10.1145/3091478.3091517}.

Various solutions attempt to solve this signal-to-noise ratio. The larger ecosystems attempt to educate their users to produce greater quality reports, while some programs will require specific items of information to gain more significant details. HackerOne has seen some success in introducing both signal requirements and rate limiter mechanisms \cite{laszka2016banishing}. The former mechanism limits those who can submit reports based on their ratio of verified to unverified submissions, whereas the latter restricts the number of reports an individual can submit in a given interval \cite{HackerOne2016Signal}. The most definitive solution to this issue is the suggestion to request appropriate features from those outlined in the 2018 update of the ISO 29147 \cite{ISO2018}. While some submissions, and platforms, may not warrant all outlined features, the inclusion of several may lead to an increase in the reproducibility of the reported vulnerability. Existing research recognises the critical role played by the reproducibility of a bug bounty submission, even suggesting that reports contain features which appear in the ISO standard \cite{Usenix2018}.

An associated problem concerns the issues surrounding duplicate submissions, a consequence of the first-come first-served attitude to verified submissions \cite{zhao2017devising,laszka2018rules}. As discussed, the race to submit first often leads to reports lacking essential details. In this scenario, a vendor or platform require further communication with the hacker to gain the necessary details. In this time,  another hacker may submit a more significant detailed report for the same vulnerability. The second report, although possibly more beneficial to the vendor, under the first-come, first-served attitude is a duplicate. The treatment of duplicates varies across bug bounty platforms. Some platforms do not monetarily reward the duplicate submitter and instead compensate with non-monetary awards. This stance towards duplicate reports rewards actively discourages detailed submissions. How bug bounty programs handle duplicate reports is in stark contrast to the closely associated software bug tracking methods, where reproducing of reports is encouraged \cite{just2008towards}.

A seminal study in this area conducted by Mallart et al. \cite{Maillart2017Given} identified a possible behavioural trend reminiscent of the St. Petersburg Paradox within bug bounty program ecosystems. The probability of finding a given number of bugs and their associated payoffs govern a hacker's aim to increase earnings. As the average bounty per program scales super-linearly, while the probability of bug discovery decays rapidly, there is less incentive to explore the depth of a vendor for bugs. Consequently, through incentivisation of increased earnings, hackers diversify their portfolio by switching programs \cite{Zhao2014Exploratory,Zhao2015Empirical}. Without greater incentivisation to outweigh the decreased probability to discover bugs, there is a potential problem with incomplete coverage possibly leading to a false perception of security. The heterogeneity of hacker skillsets may mitigate this problem to some extent \cite{hata2017understanding}. 

The issue of hackers switching programs is amplified upon the launch of a newly launched program. In this scenario, existing bounty programs will receive fewer reports, more so when rewards provided by new programs are higher.  However, the impact of this also depends on factors such as reputation and reward payouts \cite{Maillart2017Given}. This problem likely impacts bug bounty platforms where there are many programs, rather than the organisations with their own direct programs. However, existing programs can similarly exploit this phenomenon by releasing their scope in stages \cite{O'Hare2019}.

To facilitate an improved bug bounty process, we propose an implementation featuring elements of gamification to address issues found in traditional programs.

\subsection{Gamification}

\subsubsection{Nomenclature}
Several terms have been used to describe the use of game mechanics and elements in non-game settings.  One such term used is gamification, defined as \textit{``the application of gaming mechanics to non-gaming environments"}\cite{GamificationDefinition}.

The concepts of serious games and applied games are often linked with the notion of gamification, however these are distinctly different.  Serious games are \textit{``a game in which education (in its various forms) is the primary goal, rather than entertainment"} \cite{michael2005serious}.  Applied games on the other-hand are \textit{``an  implementation  of  a  subject, inspired by and designed along a context and user-centric transfer of design concepts and qualities from the game world"} \cite{schmidt2015applied}.

Thus in the context of this paper, the term gamification has been used as the appropriate term to encompass game mechanics and elements.

\subsubsection{Applications of Gamification}

Elements of gamification can be presented in many forms and the usage of particular elements should be considered in the context of the environment in which they are placed.  Existing work acknowledges there are at least 52 possible elements of gamification elements and mechanics which can be incorporated into a variety of contexts, though this is not an exhaustive list \cite{GamifiedUK}.  Mechanics and elements can include the use of social networks, encouraging exploration, challenges, and the sharing of knowledge.  User types have also been considered and provide a link to which gamification elements might be better suited to different individuals e.g. free spirits who are motivated by autonomy and may enjoy exploration elements, and disruptors who are motivated by change and may prefer a level of anonymity \cite{GamifiedUK}.  These user types have been verified by empirical research \cite{tondello2019empirical}.

Other researchers have created a taxonomy of gamification elements linked to the level of commitment required by the user when interacting with such components \cite{robinson2013preliminary} .  Many gamification elements are considered low-level commitments and are items which should be easily understood, for example, general indicators to provide users with feedback.  Others, such as an \textit{``ambiguous path to objective"} require a higher level of commitment, particularly if there are multiple puzzles to solve to reach this point.

Gamification has been used in a number of domains, namely education.  Perhaps one of the most popular examples of gamification is the Duolingo application \cite{Duolingo}, which helps people learn a new language, and has over 300 million users to date.  The application makes use of a variety of gamification elements including badges, rewards, and trophies, in addition to its expansive learner paths \cite{Gamifiedco}.

Higher Education institutions have also made use of gamification whilst teaching students.  Delft University of Technology in the Netherlands integrated gamification into a first-year BSc module on Computer Organization, and an MSc module focusing on Cloud Computing \cite{iosup2014experience}.  Mechanics utilised included points, levels and leaderboards, and additional dynamics included the use of badges, onboarding, social engagement and methods of unlocking content.  The research found that the introduction of gamification benefitted the students, facilitated interaction, and kept students motivated throughout the semester.

To explore the area further,  researchers have performed a systematic literature review examining the use of gamification in higher education \cite{subhash2018gamified}.  The work found that there was growing support for the use of gamification in Higher Education, with Spain being a leading country in the field.  Furthermore it was found that much of the research was being conducted within the field of computing, and that \textit{``points, badges, and leaderboard were the most commonly used game elements for gamification of courses"} \cite{subhash2018gamified}.

Use of the concept is also growing in the field of cybersecurity.  In 2013, research was published outlining a possible design solution to create a cyber defence training platform \cite{amorim2013gamified}.  A gamified, modular approach would allow new threats to be added to the system on a regular basis, keeping the user engaged and ensure their skills were up-to-date.

Gamification and cybersecurity have been combined to help non-experts learn about the potential consequences of cybersecurity issues.  The development of a gamified environment has been proposed which would include elements such as leaderboards and onboarding tutorials to help senior executives at critical national infrastructure facilities decide how to invest in cyber defenses \cite{cook2016using} .  Preliminary work has also been conducted to attempt to improve security awareness in non-cybersecurity experts via the use of gamification \cite{scholefield2019gamification}.  Several gamification features were included such as a specific theme (medieval RPG theme), a timer to increase pressure, consequences (points lost for wrong answers), and a leaderboard to foster competition.  A prototype quiz application was developed to educate users about password security, and a pilot study produced positive results.

\subsubsection{Gamification and Bug Bounty Programs}
Despite the usage of gamification techniques in the field of education and cybersecurity, limited work has been conducted to examine the role of these techniques in the context of a bug bounty program.
GitHub \cite{GitHub2020} employs the use of points and badges within their bug bounty program, whereas Bugcrowd \cite{Bugcrowd2020} and HackerOne \cite{HackerOne2020} feature reputation-style systems.
However, the implementation of gamification techniques into bug bounty programs is generally limited, or non-existent in the majority of cases, for example, some programs incorporate a leaderboard, or the similar concept of a Hall of Fame \cite{votipka2018hackers,ruohonen2018bug,malladi2019bug}.

To fully incorporate elements of gamification into a bug bounty program, we propose a novel process which will address issues found within existing programs.

\section{\uppercase{Methodology}}
\label{sec:methodology}
This section looks to introduce and outline a novel bug bounty process which through design, and the use of gamification, may mitigate some problems inherent in existing processes.

\subsection{Direct Crowd-vetted Bug Bounty}

\begin{figure}[t]
\centerline{\includegraphics[width=0.4\textwidth]{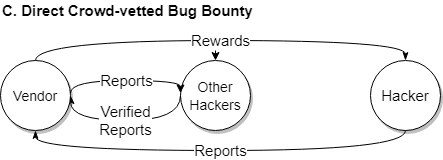}}
\caption{Proposed bug bounty process using two phases of crowdsourcing.}
\label{figBB2}
\end{figure}

The proposed bug bounty process follows a four step procedure, as presented in Figure \ref{figBB2}.

\begin{enumerate}
    \item A Hacker discovers a vulnerability and reports this finding to the vendor through appropriate channels.
    \item After optionally validating the report and checking for duplicates, the vendor redistributes this submission to a field of vetted hackers.
    \item These hackers, if possible, verify the original submission, and report back to the vendor on the verification and reproducibility of the report.
    \item The vendor will receive an actionable vulnerability report or sound reasoning to dismiss the submission. The original hacker will receive a reward or feedback depending on the success of verification, while the verifiers will receive a reward regardless.
\end{enumerate}

While this proposal utilises a direct process, an intermediary exists in the form of a vetted hacker verification. This peer-review process is the unique aspect of the methodology, introducing further crowdsourcing into the bug bounty process.

This process poses many strengths. With less vendor interaction required, this process provides a significantly reduced overhead. In theory, the further crowdsourcing process should provide an increase in reports verified as a consensus is required. Furthermore, this process shows potential as an educational resource. verifiers will become exposed to new methods when reproducing submissions; who can then hunt bugs using the same methods - similar to disclosed reports \cite{Zhao2015Empirical,votipka2018hackers}.

The proposed process is not without fault. Without vendor intervention, the verification process would dismiss a valid yet out of scope vulnerability. Although out of scope, if critical, these vulnerabilities are typically addressed on a case-by-case basis in existing bug bounty platforms. Additionally, this optional vendor validation can function as a safety-net to catch duplicate vulnerabilities with a possible alternative being an obfuscated current issue tracker. However, the extent of this vendor intervention comes at the cost of increased overhead. Furthermore, a poor uptake or sustained decrease in overall participants would significantly limit the effectiveness of a program, potentially increasing the time to respond, and may contribute to further issues and threats. Moreover, the increase in participants in the process comes with an array of potential issues. In comparison to existing methods, this process introduces an increased demand on the rewards required to operate.

Nevertheless, this process presents significant opportunities for further development. Whereas current bug bounty processes concentrate on the reporter, this novel process allows for the vendor organisation to build relationships with verifiers as well. Through these relationships, a community focused on the program can be built \cite{votipka2018hackers}. Where the more reputable individuals can gain invitations to private program-like access to new scope in creases. As there is a greater number of external participants in this process in comparison to existing methods, there are more opportunities to implement gamification elements towards participation, incentivisation and rewards.

However, this process does introduce specific threats arising from the introduction of individuals upstream other than the vendor \cite{hata2017understanding}. One particular issue is the difficulty in maintaining a consistent barometer for verified vulnerabilities with a rotating cast of verifiers of differing expertise \cite{sana2018crowd}. These verifiers themselves also introduce several potential threats; a verifier may breach confidentiality and leak information regarding the vulnerability resulting in external exploitation and concomitant issues, including negative press and stock fluctuation. Similarly, verifiers may collude with the reporter to receive rewards without proper scrutiny. On the other hand, a verifier may seek the dismissal of a report, only to then submit the same issue themselves.

The proposed methodology borrows heavily from work by Su and Pan \cite{Su2016Crowdsourcing}. The original proposal appears to be the first to suggest the crowdsourcing of both vulnerability discovery and verification. However, this work differs from the original by limiting the number of roles in the bug bounty process and allowing individuals to verify vulnerabilities as they see fit. Additionally, the original proposal appears to focus heavily on discovery and verification, neglecting other procedures in the bug bounty process.  

\subsection{Aspects of Gamification}
When developing a gamified bug bounty program, it is vital to consider challenges typically found in traditional bug bounty programs; the aspects of gamification should mitigate these issues, and provide an improved user experience.  Following on from section \ref{bugbountyissues}, gamification elements implemented should aim to address several key areas:

\begin{itemize}
    \item \textbf{A1:} Ensure the inclusion of all elements of the overall bug bounty process for each user, i.e. discovery and verification, and consider the individuals' expertise in each of these areas. 
    \item \textbf{A2:} Implementation of gamification elements should enhance and support the bug bounty experience.
    \item \textbf{A3:} Incentivise any and all participation in the program.
    \item \textbf{A4:} Encourage prolonged activity across various aspects of the program. 
\end{itemize}

Existing bug bounty programs typically give little consideration to such features, however these areas deserve greater examination within this context.

Gamification elements should encompass all possible steps of the bug bounty process to satisfy the criteria outlined in A1. Ensuring symmetry in gamification elements between vulnerability discovery and verification would be an example. This symmetry would value both processes equally, potentially mitigating the possible neglect for one system over another. However, such systems should be dynamic and tailored to one's strengths and weaknesses to provide an engaging experience.

Some bug bounty programs currently incentivise the higher quality submissions, thus enhancing the bug bounty experience and effectiveness, through additional monetary rewards \cite{zhao2017devising}. Gamification elements could substitute or reinforce such a process. Encouraging well-mannered correspondence, promoting diverse vulnerability discovery methods, and alignment to pressing business objectives are all goals further gamification implementations can pursue in the name of program effectiveness. 

Bug bounty participants are heterogeneous, with the majority never achieving a successful submission \cite{hata2017understanding}. However, in the proposed method, these less successful participants can still contribute to the program through the verification process. From which they can learn from peers, and apply this new knowledge in their pursuit of vulnerabilities. In this scenario, a prolific verifier encounters significant incentivisation to attempt vulnerability discovery and vice-versa. Other methods to encourage participation could include initial higher incentives and rewards overall. 

While scope increases and significant codebase updates can sustain participation in the bug bounty process, A4 strives to achieve this through gamification aspects. With a significant decrease in activity typical after the initial launch window of a bug bounty, program managers will attempt to repeat the success through attracting participants to return with limited-time events with greater rewards. Such events often coincide with scope increases. Providing solely monetary incentivisation during this event can lead to a significant signal-to-noise ratio impacting the effectiveness of this strategy \cite{10.1145/3091478.3091517}. Nevertheless, this strategy could part or whole substitute monetary incentivisation for gamification elements. Additionally, elements can encourage continued activity through incentivisation of repetitive actions in a given time frame, i.e. verifying a sum of vulnerabilities in a given week. 

To address these, we seek to include gamification concepts which may appeal to all users of the system (named as general and rewards elements), and more specific elements which have a a link to user types (Table \ref{tab:gamificationelements}).  In exploring which aspects of gamification which may be suited to a bug bounty program, Gamification Inspiration Cards were used \cite{GamifiedUK}, in addition to considering the context of such a program.

\begin{itemize}
\item \textbf{General: }These gamification concepts can work for a variety of user types and scenarios, and can be implemented in a number of ways.
    \begin{itemize}
    \item \textit{Investment}- it is vital to ensure users are invested in the system (purpose) and that they will continue to engage.  A variety of methods can be used to foster investment in the system, including rewards such as badges and points, leaderboards, and challenges, whereby the scope increases to keep more experienced hackers engaged.

    \item \textit{Progress and feedback- } Users of the system need to receive feedback in order to remain engaged, and to highlight their advancement through the system.  Feedback can be provided in a number of forms, including badges and achievements.
    \item \textit{Onboarding and tutorials- }If users are unsure how the platform works, this could lead to a lack of engagement after a short time using the system.  The will help users get used to the aspects of the system and help engagement.
    \end{itemize}

\item \textbf{Rewards: }
    \begin{itemize}
    \item \textit{Fixed reward schedule- } These rewards can be awarded based on specific events within the system, for example, the first submission to the program could provide the user with double the number of points they would typically receive.  These rewards could also be delivered based upon milestone achievements, such as awarding a badge for every 10 verified submissions to the program.  Providing fixed objectives may assist in addressing issues concerning sustaining participants and their attention.
    
    \item \textit{Other rewards- }  There are other rewards which can be included within the program however these are typically tied to user type (e.g. the Player user type is said to be motivated by rewards).  Rewards pertaining to specific user types will be discussed in Table \ref{tab:gamificationelements}.
    
    \end{itemize}
\end{itemize}

The culmination of these gamification elements will ensure that the system is not relying on simplistic leaderboards.

There are a number of user types and associated gamification elements which are deemed appropriate for use in the bug bounty program \cite{GamifiedUK}.  These include:

\begin{itemize}
    \item \textbf{Players: }Users who are motivated by rewards.  This links in with work conducted by Hata et al. \cite{hata2017understanding} \textit{``Archetype 2"} (project specific contributor to a bug bounty program) and \textit{``Archetype 3"} (non-project specific contributor to a bug bounty program) both tend to be motivated by rewards, with \textit{``Archetype 2"} spending more effort than \textit{``Archetype 3"}\cite{hata2017understanding}.
    \item \textbf{Achievers: }Those who seek to master a particular subject area.
    \item \textbf{Socialisers: }Individuals who want some form of social interaction from the program.
    \item \textbf{Philanthropists: }Individuals who are altruistic and require meaning or a purpose behind a platform to engage with it.  This user type links in with existing research which defined \textit{``Archetype 2"}, a type of contributor to a bug bounty program who is motivated to help users and to make the end product secure \cite{hata2017understanding}
\end{itemize}

\begin{table*}[]
\centering
\begin{tabular}{|p{2cm}|p{1.6cm}|p{3cm}|p{5cm}|p{2cm}|}
\hline
\rowcolor[HTML]{C0C0C0} 
\textbf{User Type} & \textbf{User Motivation} & \textbf{Gamification Element} & \textbf{Description \& Implementation} & \textbf{Issue(s) Addressed} \\ \hline
 &  & Badges \& Achievements & Feedback to highlight the user's progression through the system. & A1, A2, A3, A4 \\ \cline{3-5} 
 &  & Leaderboards & Compares users of the system to each other, and helps to foster competition.  The bug bounty program should be transparent to allow these users to link back to the bugs found and reports verified. & A1, A3, A4 \\ \cline{3-5} 
\multirow{-3}{*}{Players} & \multirow{-3}{*}{Rewards} & Points \& Experience & A form of feedback to show users they are making progress and to elevate their position on the leaderboard. & A1, A2, A3 \\ \hline
 &  & Certificates & Symbolises mastery and could be used to reward more advanced hackers within the system. & A1, A2, A3 \\ \cline{3-5} 
\multirow{-2}{*}{Achievers} & \multirow{-2}{*}{Mastery} & Challenges & Increased scope to cater to advanced hackers.  For these individuals it ensures they are challenged when using the program, leading to incentive for continued engagement. & A1, A2, A3, A4 \\ \hline
 &  & Social Status & Opportunities for new relationships e.g. joining a well-performing group or team.  Also fosters a sense of belonging and may enhance purpose & A1, A2, A3 \\ \cline{3-5} 
 &  & Competition & The nature of a bug bounty program lends itself to competition i.e. hackers are competing to be the first person/group to find a vulnerability, and to document it. & A2, A3 \\ \cline{3-5} 
\multirow{-3}{*}{Socialisers} & \multirow{-3}{*}{Relatedness} & Guilds or Teams & This would allow a user to form a group or team (either on a local friends group level, or as part of a massive online team). & A2, A3 \\ \hline
Philanthropists & Purpose & Meaning or Purpose & Clear information on what the program is trying to achieve, and how the user is central to the goal. & A2, A3 \\ \hline
\end{tabular}
\caption{\label{tab:gamificationelements}Suitable aspects of gamification for the bug bounty program.} 
\end{table*}

\subsection{Possible Implementation}

With the potential for trust-related issues, operating the proposed bug bounty process in an organisation with no existing relationship to the external hackers may invite unwanted risk \cite{Nicebugs2020}. One of the organisations with an existing relationship to an array of hackers is a tertiary educational institution, where hackers in question are students. In this scenario, any significant trust issues equate to disciplinary issues. While in comparison to existing methods, this process seems most appropriate to a university setting.

Firstly, the reduced overhead takes away a considerable upfront burden from the information services department. However, the bug bounty could result in numerous sophisticated bug reports beyond the technical knowledge and remit of the services department. Secondly, the compatibility with gamification provides an economical incentivisation measure. Lastly, the verification process provides an extracurricular educational resource; however, it relies on a dedicated uptake to do so.

Implementation of this process at a university possibly aligns to the Academic Centres of Excellence in Cyber Security Education (ACE-CSE) criteria set forth by the UK's National Cyber Security Centre \cite{NCSC2020ACE}. The ACE-CSE program attempts to offer institutions means to demonstrate a systematic commitment to cybersecurity. Outwith academic responsibilities, the applicant institution must evidence cybersecurity faculty possessing a role in ensuring the cybersecurity of the university's networked infrastructure. Faculty cooperating with the operation, administration, and encouraging participation in such a bug bounty program may satisfy this criterion.

The operation of bug bounty programs at education institutions seems limited to select American universities, yet, some British counterparts do operate vulnerability disclosure policies \cite{York2020,Oxford2020,Bourne2020}.

\section{\uppercase{Discussion}}
\label{sec:discussion}
Research has noted that gamification can improve student-engagement when used in Higher Education, in addition to increasing motivation and performance\cite{subhash2018gamified}.  Thus is it plausible that a bug bounty program could benefit from an application of gamification components.

Commonly used gamification elements found in Higher Education (particularly in computing subjects) include points, badges, and leaderboards, however, it should be stressed that gamification goes beyond these elements \cite{GamifiedUKMorePoints}.  Work by Fischer et al. \cite{fischer2016gamifying} highlights that the implementation of gamification must fit within the scope of the technologies and applications used, and emphasise that emotions, engagement, and motivation must also factor in to the design of any such gamification system.

Although gamification appears to be widely used in Higher Education, commonly in computing-based subjects, it has its limitations \cite{subhash2018gamified}.  Previous work has implemented a gamification strategy in a course teaching introductory C programming at undergraduate level \cite{ibanez2014gamification}.  To achieve this, a number of gamified activities were developed, making use of points and badges, generally improving knowledge acquisition.  Gamification presented issues for some students: some became disengaged when reaching the round number of one hundred points, choosing not to continue with additional tasks.  Though the proposed bug bounty implementation seeks to mitigate such issues, it is nonetheless a risk that users may become disengaged after a period of time.

Prior studies that have noted the importance of reward evolution as the program and organisation security matures \cite{Zhao2015Empirical,al2018friendly}. One study suggests front-loading the launch of a program with high payouts, before adjusting payout structure to decrease payout towards a market average to allow for greater rewards for complex vulnerabilities \cite{votipka2018hackers}. By doing so, a program can incentivise participation during the initial release and continue to encourage dedicated participation when the number of successful reports tapers. However, such a strategy may be less effective when a bug bounty operates on a significantly limited budget or relies on non-monetary rewards. Therefore, those in this scenario may require alternative solutions to keep encouraging participation. 

While the effectiveness of the proposed bug bounty program remains speculative, the contrast with existing methods may lack significance without a robust remediation process \cite{Nicebugs2020}. However, without either paying a platform for verification or crowdsourcing via the proposed method, the team responsible for the remediation process may have to screen a high signal-to-noise ratio, performing verification and corresponding with hackers where necessary. By offloading these responsibilities to a bug bounty platform or the crowd themselves, the remediation process can receive greater focus from the internal security team. Nevertheless, the resolution of any issues and ineffectiveness with the remediation process is essential before the implementation of a bug bounty process.

A natural progression of this work is required to establish the viability of this novel process in practice. These future investigations may wish to explore implementations beyond the educational sector. Additionally, this paper shows there is ample room for further development and implementation of gamification in the bug bounty domain.

\section{\uppercase{Conclusions}}
\label{sec:conclusion}
By leveraging gamification techniques against common bug bounty issues, this paper proposed a new bug bounty implementation intending to provide a crowdsourcing cybersecurity solution and an educational resource. Using a higher education institution as a potential use case allows for the consideration of the application of gamification aspects to improve program effectiveness. Without a trial implementation and concomitant study, the reality of the process remains uncertain. However, this proposal highlights the limited use of gamification in bug bounties thus far and contributes approaches to implement aspects which may increase program effectiveness.

\bibliographystyle{apalike}
{\small
\bibliography{references}}

\end{document}